\documentclass[useAMS,usenatbib]{mnras}

\usepackage{graphicx}
\usepackage{grffile}\usepackage{epsf}
\usepackage{amsmath}
\usepackage{float}
\usepackage{fixltx2e}
\usepackage{bm}

\newcommand{\kms}{{\rm km}\,{\rm s}^{-1}}

\newcommand{\msolar}{{\rm M}_{\odot}}

\newcommand{\gad}{{\sc Gadget-3}}

\newcommand{\OVI}{\hbox{O\,{\sc vi}}}
\newcommand{\OVII}{\hbox{O\,{\sc vii}}}
\newcommand{\OVIII}{\hbox{O\,{\sc viii}}}

\newcommand{\HI}{{\hbox{H\,{\sc i}}}}

\begin{document}
\title[Deviations from HSE in the CGM]{Deviations from hydrostatic equilibrium in the circumgalactic medium: spinning hot haloes and accelerating flows}

\author[B. D. Oppenheimer]{
\parbox[t]{\textwidth}{\vspace{-1cm}
Benjamin D. Oppenheimer$^{1}$\thanks{benjamin.oppenheimer@colorado.edu}\\\\
$^1$CASA, Department of Astrophysical and Planetary Sciences, University of Colorado, 389 UCB, Boulder, CO 80309, USA
}}
\maketitle

\pubyear{2017}

\maketitle

\label{firstpage}

\begin{abstract}

  Hydrostatic equilibrium (HSE), where the thermal pressure gradient
  balances the force of gravity, is tested across a range of simulated
  EAGLE haloes from Milky Way $L^*$ haloes ($M_{200}\approx 10^{12}
  \msolar$) to cluster scales.  Clusters ($M_{200}\ga 10^{14}
  \msolar$) reproduce previous results with thermal pressure
  responsible for $\sim 90\%$ of the support against gravity, but this
  fraction drops for group-sized haloes ($M_{200}\approx 10^{13}
  \msolar$) and is even lower ($40-70\%$) for $L^*$ haloes between
  $0.1-0.3 R_{200}$.  Energy from feedback grows relative to the
  binding energy of a halo toward lower mass resulting in greater
  deviations from HSE.  Tangential motions comprise the largest
  deviation from HSE in $L^*$ haloes indicating that the hot
  circumgalactic medium (CGM) has significant sub-centrifugal rotation
  and angular momentum spin parameters $2-3\times$ higher than the
  dark matter spin parameters.  Thermal feedback can buoyantly rise to
  the outer CGM of $M_{200}\la 10^{12} \msolar$ haloes, both moving
  baryons beyond $R_{200}$ and feeding uncorrelated tangential
  motions.  The resulting hot halo density and rotation profiles show
  promising agreement with X-ray observations of the inner Milky Way
  halo, and we discuss future observational prospects to detect
  spinning hot haloes around other galaxies.  Acceleration and radial
  streaming motions also comprise significant deviations from HSE,
  especially net outward accelerations seen in $L^*$ and group haloes
  indicating active feedback.  Black hole feedback acts in a
  preventative manner during the later growth of group haloes,
  applying significant accelerations via shocks that do not feed
  tangential motions.  We argue that HSE is a poor assumption for the
  CGM, especially in the inner regions, and rotating baryonic hot
  haloes are a critical consideration for analytic models of the CGM.

\end{abstract}


\section{Introduction}  

The efforts to reveal the circumgalactic gas reservoirs that supply
galaxies with the fuel to create stars have made significant steps in
recent years with instruments including the Cosmic Origins
Spectrograph (COS) on {\it Hubble} and X-ray instruments on {\it
  Chandra} and {\it XMM}.  UV absorption line modelling has accounted
for a significant cool ($\sim 10^4-10^5$ K) gas reservoir extending at
least 150 kpc and out to the virial radius \citep[$R_{200}$,
  e.g.][]{wer14,borth16,kee17}.  The ambient halo medium that theory
predicts should be heated to the virial temperature ($T_{\rm vir}$),
which is $\sim 10^6$ K for a $10^{12} \msolar$ ``$L^*$'' halo, remains
a challenge to detect out to the virial radius, even though there are
detections in the inner CGM of luminous spirals and ellipticals
\citep[e.g.][]{and11, li13} and the Milky Way halo
\citep[e.g.][]{mil15}.

These efforts, although admirable, indicate just how nascent the
accounting of the CGM multiphase mass budget is \citep[][and
  references therein]{tum17}.  Yet, the main reason we study the CGM
is to understand the dynamics of gas fueling, ejection, and recycling
that regulate galaxy formation and evolution.  The gulf between
observations that are only beginning to constrain mass budgets and the
theory of how circumgalactic gas feeds star formation is vast.
Fundamental theoretical questions are rarely broached including-- is
the CGM in hydrostatic equilibrium (HSE)?; what comprises the
deviations from HSE?; and what are the implications of such
deviations?

Studies at the cluster scale have approached these questions, because
emission can be measured to $R_{500}$ (the radius that encloses an
overdensity of $500\times$ the critical overdensity) and beyond
\citep[see compilation of observational results by ][]{mcc17}.
Additionally, hydrostatic mass estimates of clusters are used as
cosmological tools.  A number of studies have attempted to quantify
deviations from HSE using cosmological hydrodynamic simulations of
clusters, usually finding deviations of 10-20\% from HSE, but often
with differing conclusions as to the specific cause \citep{fan09,
  lau09, sut13, nel14, bif16}.  Most of these studies quantify the
hydrodynamics of intra-cluster medium (ICM) using the Euler equation
of momentum conservation,

\begin{equation}
  \frac{d{\bm v}}{dt} = -\nabla \Phi - \frac{1}{\rho_{\rm gas}} \nabla P,
\end{equation}

\noindent where $\Phi$ is the gravitational potential, and ${\bm v}$,
$\rho_{\rm gas}$, and $P$ are the velocity, density, and pressure of
the gas.  HSE applies if $\frac{d{\bm v}}{dt}=0$.

Halo gas is often assumed to be in HSE in analytically-based models of
the CGM \citep[e.g.][]{mal04,tep15,fae17,mat17}, and rarely is HSE
tested in simulations with the exception of the \citet{fie17}
idealized cases.  Quantifying HSE is more fraught with difficulty for
the CGM than the ICM, because 1) the CGM is definitely multiphase
\citep[e.g.][]{wer16}, while the ICM is dominated by hot gas
\citep[e.g.][]{gon13}, and 2) deviations are likely larger owing to
the lower binding energy of these haloes, allowing feedback from star
formation and active galactic nuclei (AGN) to create greater
disturbances.  Nevertheless, the hydrodynamic state of the CGM should
be a pre-requisite to modelling the formation, dynamics, and fate of
the cool and hot gas traced in UV absorption line spectroscopy and
X-ray emission and absorption probes.

We use EAGLE (Evolution and Assembly of GaLaxies and their
Environments) simulations \citep{sch15,cra15,mca16} to quantify the
hydrodynamic state in haloes hosting $L^*$ galaxies that are actively
star-forming ($M_{200}\approx 10^{12} \msolar$) and ``group''-sized
haloes hosting mainly passive galaxies ($M_{200}\approx 10^{13}
\msolar$).  Our main simulations are a set of EAGLE zoom haloes
that have been tested for a number of CGM studies, including $\OVI$
\citep{opp16} and low metal ions \citep{opp17b}.  We are able to
output the acceleration vector in the zooms, which can be significant
for dynamics.

We lay out the theoretical method to deconstruct the Euler equation
into halo support terms, which determine deviations from HSE, and
introduce our suite of simulations in \S\ref{sec:setup}.  We test our
method on three haloes spanning the hot halo regime: a cluster, a
group, and a Milky Way-mass $L^*$ halo.  We present the main results
from our samples in \S\ref{sec:gentrend} where we focus on normalized
halo quantities allowing cross-sample comparisons.  These results
include the fractional deviations from HSE, which leads us into a
discussion of velocities normalized to the virial velocity of the
halo.  Masses and angular momentum spin parameters subdivided into
baryonic and dark matter components, as well as reservoirs of baryonic
energies, continue our exploration.

Our discussion begins in \S\ref{sec:selfsim} asking how the
self-similar scaling relations expected from dark matter structure are
broken using the perspective of the CGM.  We advocate that feedback
rather than cooling causes the fundamental deviations from HSE, and
consider observations that can determine the primary deviation from
HSE in $L^*$ haloes-- gas with significant rotational and tangential
motion in \S\ref{sec:obs}.  Future directions for observations,
analytical models, and simulations are discussed in
\S\ref{sec:future}.  We summarize in \S\ref{sec:summary}.  All results
are at $z=0$, unless otherwise noted.

\section{Theoretical setup} \label{sec:setup}

\subsection{Euler equation and terms}

We begin with the Euler equation:

\begin{equation}
\frac{\partial {\bm v}}{\partial t} + ({\bm v} \cdot \nabla) {\bm v} = -\frac{1}{\rho_{\rm gas}} \nabla P - \nabla \Phi.
\end{equation}

\noindent
The gravitational potential in the last term relates to the total mass
(dark matter (DM) plus gas plus stars) within volume $V$, via the
Poisson equation integrated at a surface $\partial V$ encompassing
$V$ using Gauss's Law,

\begin{equation}
M_{\rm tot} =  \frac{1}{4\pi G} \oint_{\partial V} d{\bm S} \cdot \nabla \Phi,
\end{equation}

\noindent where $d{\bm S}$ is the surface element.  We then apply
Gauss's law to the Euler equation, putting all forces balancing
gravity on the right-hand side:

\begin{equation}\label{equ:Euler_mass}
M_{\rm tot} = \frac{1}{4\pi G}  \oint_{\partial V}  d{\bm S} \cdot  \big( -\frac{1}{\rho_{\rm gas}} \nabla P -  ({\bm v} \cdot \nabla) {\bm v} - \frac{\partial {\bm v}}{\partial t}  \big).
\end{equation}

\noindent as done in previous cluster studies.

Adopting a spherical surface, we decompose the right-hand side
of \ref{equ:Euler_mass} into four effective mass terms:

\begin{equation} \label{equ:effmass}
  M_{\rm tot} = M_{\rm therm} + M_{\rm rot} + M_{\rm stream} + M_{\rm acc}
\end{equation}

\noindent where  

\begin{equation} \label{equ:masstherm}
M_{\rm therm} = -\frac{1}{4\pi G} \oint_{\partial V} dS \frac{1}{\rho_{\rm gas}} \frac{\partial P}{\partial R},
\end{equation}

\begin{equation} \label{equ:massrot}
M_{\rm rot} = \frac{1}{4\pi G} \oint_{\partial V} dS \frac{v_\theta^2 + v_\phi^2}{r},
\end{equation}

\begin{equation} \label{equ:massstream}
M_{\rm stream} = - \frac{1}{4\pi G} \oint_{\partial V} dS \big( v_r \frac{\partial v_r}{\partial R} + \frac{v_\theta}{R} \frac{\partial v_r}{\partial \theta} + \frac{v_\phi}{R \sin \theta} \frac{\partial v_r}{\partial \phi} \big),
\end{equation}

\noindent and 

\begin{equation} \label{equ:massacc}
M_{\rm acc} = -  \frac{1}{4\pi G} \oint_{\partial V} dS \frac{\partial v_r}{\partial t}.
\end{equation}
\noindent in spherical coordinates, $R$, $\theta$, and $\phi$. 

The first term in Equation \ref{equ:effmass}, $M_{\rm therm}$,
represents the thermal pressure gradient of the gas, and should equal
$M_{\rm tot}$ in thermal HSE.  At the other extreme, gas that is
isobaric at a surface will have no thermal support, and the remaining
terms should sum up to $M_{\rm tot}$.  Positive pressure gradients can
result in negative $M_{\rm therm}$ terms, but this is rare integrating
across a spherical surface around a halo.

The second and third terms, $M_{\rm rot}$ and $M_{\rm stream}$, are the
inertial terms, derived from the $({\bm v} \cdot \nabla) {\bm v}$ term
in the Euler equation.  We refer to $M_{\rm rot}$ as ``tangential''
support, since it contains both mean and random tangential motions.
Mean tangential motions are more related to what we refer to as
correlated ``rotational'' or centrifugal support.  The rest of the
$M_{\rm rot}$ term arises from uncorrelated motions in the tangential
direction, which we show can be significant.  This term is always
positive by definition.

Like $M_{\rm rot}$, the streaming term encompasses radial motions both
correlated and random.  Correlated motions include 1) infalling gas
that slows down at lower radii resulting in a negative $M_{\rm
  stream}$, and 2) outflowing gas that slows down at larger radii
yielding a positive $M_{\rm stream}$.  This term also includes
uncorrelated, random dispersion in the radial direction above the grid
resolution on which we calculate these quantities (see
\S\ref{sec:sim}), and becomes positive in that case.  This latter term
can be considered turbulent pressure above the grid resolution.  

The acceleration term, $M_{\rm acc}$, indicates temporal variations in
the radial gas velocities, and becomes positive for gas accelerating
toward the halo center, and negative for gas accelerating away from
the halo center.  For example, if $M_{\rm therm} < M_{\rm tot}$ at a
surface, and the inertial terms are zero, the gas is not pressure
supported and will accelerate inward, such that $M_{\rm therm} +
M_{\rm acc} = M_{\rm tot}$.

\citet{lau13} clarifies the meaning of $M_{\rm rot}$ and $M_{\rm
  stream}$ calculated using Euler ``summation'' terms as we and
\citet{sut13} do.  Other studies \citep[e.g.][]{nel14} use the
``average'' terms that define ``$M_{\rm rot}$'' and ``$M_{\rm
  stream}$'' as correlated tangential and radial motions, and then a
$M_{\rm rand}$ to quantify random motions in the radial and tangential
directions.  Our $M_{\rm rot}$ and $M_{\rm stream}$ as defined above
include correlated and random motions, but we break down these two
types of motions when discussing velocities in \S\ref{sec:vels}.

We discuss other sources of pressure in \S\ref{sec:future} not
included in our simulation, such as cosmic ray and magnetic pressure,
which we argue are less important for hot, $T>10^5$ K gas than for
cool, $T\sim 10^4$ clouds.  Small-scale turbulent pressure below our
grid resolution is tested for hot gas, and we find this not to be a
significant contributor to the Euler mass terms.

Because we are considering the contributions of Euler terms to the
{\it support} against gravity, we define the normalized variable
\begin{equation}
{\mathcal S}_{term}(R) \equiv \frac{M_{term}(R)}{M_{\rm tot}(R)}.
  \end{equation}
\noindent These normalized support variables better highlight the
fractional contributions to and deviations from HSE.  Additionally,
since we cannot ``weigh'' CGM haloes with current observational
probes, effective mass terms have less meaning for the CGM.

\subsection{Simulations} \label{sec:sim}

We apply the above analysis to the EAGLE zoom simulations introduced
in \citet{opp16} as our primary simulation set in this study.  We
refer the reader to \S2 of \citet{opp16} for further details, but
briefly describe the simulations here.  The EAGLE code \citep{sch15,
  cra15} is a heavily modified version of the N-body+Smoothed Particle
Hydrodynamic (SPH) \gad~code previously described in \citet{spr05}.
EAGLE uses a pressure-entropy SPH formulation along with several other
modification referred to as {\tt Anarchy} SPH \citep{schal15}.
Subgrid prescriptions for radiative cooling \citep{wie09a}, star
formation \citep{sch08}, stellar evolution and chemical enrichment
\citep{wie09b}, and superwind feedback associated with star formation
\citep{dal12} and black hole (BH) growth \citep{ros15}, are included.
\citet{pla14} cosmological parameters are assumed.

The stellar and BH feedback are both thermal in nature.  The stellar
feedback heats gas particles by $\Delta T = 10^{7.5}$ K 30 Myr after a
star particle forms.  The BH feedback adds $\Delta T = 10^{9.0}$ K in
the ``Recal'' prescription we use for most of the runs and $\Delta T
=10^{8.5}$ K in the ``Ref'' prescription for the cluster sample
described below.  The BH efficiencies, discussed in
\S\ref{sec:energies}, are unchanged between the prescriptions, meaning
the latter prescription has a higher mass loading factor.

We use 3 main samples containing 9 haloes each throughout: ``$L^*$''
($M_{200}\approx 10^{12} \msolar$), ``group'' ($M_{200}\approx 10^{13}
\msolar$), and ``cluster'' ($M_{200}\approx 10^{14} \msolar$).  The first
two samples are zooms introduced in \citet{opp16}, and the latter is
selected from the 100 Mpc EAGLE volume.  The zooms use the Recal
prescription and are run $8\times$ the EAGLE fiducial resolution,
equivalent to the EAGLE-HiRes simulation volume.  The zooms were run
at low redshift with a non-equilibrium ionization and cooling module
\citep{opp13a, ric14a} following 136 ionization states for 11
elements.  \citet{opp16, opp17b} showed that the non-equilibrium
module insignificantly alters the dynamics of the gas under a uniform
extragalactic ionization background, which makes them representative
of the EAGLE Recal prescription.

The 9 $L^*$ zooms, listed as Gal001-Gal009 in Table 1 of
\citet{opp16}, have a $z=0$ halo mass range of
$M_{200}=10^{11.85}-10^{12.28} \msolar$ with a median mass
$10^{12.02} \msolar$.  The 9 group zooms, listed as Grp001-Grp009 in
the same table, have $M_{200}=10^{12.76}-10^{13.21} \msolar$ and a
median mass of $10^{12.96} \msolar$ at $z=0$.  The $L^*$ central
galaxies have stellar masses,
log[$M_*/\msolar$]$=10.13^{+0.18}_{-0.10}$, and specific star
formation rates, defined as star formation rate divided by $M_*$,
log[sSFR/${\rm yr}^{-1}$]$=-10.01^{+0.10}_{-0.19}$.  The respective
values for the group centrals are $10.76^{+0.11}_{-0.09}$ and
$-10.91^{+0.39}_{-0.77}$, which corresponds to 4 of 9 galaxies below
the passive threshold of sSFR$=10^{-11.0} {\rm yr}^{-1}$ used by
\citet{sch15} indicating that there are low levels of star formation
in most group galaxies.

We use the nomenclature {\it M}[log($m_{\rm SPH}/\msolar$)], where
$m_{\rm SPH}$ is the initial mass of SPH particles to indicate
simulation resolution.  {\it M5.3}, signifying
$m_{\rm SPH}=2.3\times 10^5 \msolar$, is our fiducial resolution,
which is the same resolution as the parent EAGLE-HiRes Recal-L025N0752
volume, where L025N0752 is the EAGLE nomenclature indicating box size,
25 comoving Mpc, and number of SPH and DM particles on a side, 752.
The $L^*$ zooms were selected from the Recal-L025N0752 volume, and the
group zooms were selected from the EAGLE Ref-L100N1504 volume.  Both
sets of haloes were selected to host typical galaxies in the evolved
Universe.  They were run at {\it M5.3} resolution using the Recal
prescription beginning from $z=127$ initial conditions.  We output the
acceleration vector for several snapshots between $z=0.05$ and $0$.

We also select a ``cluster'' sample of $\sim 10^{14} \msolar$ haloes
from the EAGLE Ref-L100N1504 box to test how well we reproduce
previous findings, and to provide reference for our lower mass CGM
zooms.  These 9 haloes range between $M_{200}=10^{13.88}-10^{14.49}
\msolar$ with a median mass of $10^{13.98} \msolar$ at $z=0$, and are
the 9 most relaxed of the 12 most massive haloes in the 100 Mpc EAGLE
volume.  The clusters have the native {\it M6.2} resolution of the
Ref-L100N1504 volume, signifying $m_{\rm SPH}=1.8\times10^6 \msolar$.
The central galaxies have values of
log[$M_*/\msolar$]$=11.52^{+0.10}_{-0.06}$ and log[sSFR/${\rm
    yr}^{-1}$]$=-11.23^{+0.82}_{-0.41}$.

Finally, we take the 5 most massive $L^*$ haloes, and make a Milky Way
(MW) halo sample with $M_{200}=10^{12.02}-10^{12.28} \msolar$ and a
median mass of $10^{12.17} \msolar$.  We rerun these 5 haloes using a
``no-wind'' prescription at {\it M5.3} resolution, where no stellar
and BH feedback occurs.  We select these same 5 haloes from the
Ref-L025N0376 volume to make a sample of {\it M6.2} $L^*$ zooms to
test numerical convergence.\footnote{We also test our method on the
  haloes of the EAGLE Ref-L100N1504 box, and present results on our
  website: http://www.colorado.edu/casa/hydrohalos.}

We calculate the effective mass terms in Equations
\ref{equ:masstherm}-\ref{equ:massacc} by gridding our simulations onto
spherical coordinate systems centered on a halo's potential minimum in
space and velocity.  We use 72 logarithmically spaced radial
bins\footnote{Linear radial spacing was tested and gives similar
  results, but logarithmic spacing better samples the scales of the
  CGM.} spanning $0.032-2\times R_{200}$ and 180 angular bins per
radius (10 $\theta$ bins and 18 $\phi$ bins).  The cluster sample uses
half as many bins.  $M_{\rm tot}(R)$ is computed by summing all gas,
star, and DM particles inside $R$.  The other terms are calculated as
a function of $R$ by summing up the grid points across the spherical
surface.  Partial derivatives of pressure and radial velocity for the
$M_{\rm therm}(R)$ (Equ. \ref{equ:masstherm}) and $M_{\rm stream}(R)$
(Equ. \ref{equ:massstream}) respectively are calculated across
adjacent spherical grid cells.

We tested a number of different ways to sum the Euler terms using
differently defined grid cells, and we chose to grid {\it all} gas
particles and not just the {\it hot}, $>10^5$ K gas particles that
provide thermal support for the halo.  The CGM is multiphase, and our
method is not perfectly suited for cool clouds embedded in a hot
medium.  However, we find converged results for the non-thermal Euler
terms when summing either all gas or just hot gas using different
sized radial bins using both logarithmic and linear spacing, which
makes our results for ${\mathcal S}_{\rm rot}$, ${\mathcal S}_{\rm
  stream}$, and ${\mathcal S}_{\rm acc}$ robust with grid cell
definition. The biggest variation is in ${\mathcal S}_{\rm therm}$,
but we feel confident that we understand why this is.  We tested
summing in grid shells with no angular coordinates and found that
${\mathcal S}_{\rm therm}$ using only hot particles provides a similar
answer as calculating ${\mathcal S}_{\rm therm}$ using all particles
summed in angular grid cells.  Cool gas associated either with
filaments, satellite galaxies, or CGM clouds biases the thermal
pressure gradient, but they are confined to a relatively small number
of angular cells per radius due to their small filling factor, thereby
allowing the summation of the pressure gradient to be determined
mainly by hot gas.  An additional test of using only the grid cells
with $>90\%$ hot gas also yielded similar results.  These tests also
informed our choice for using relatively fine radial spacing of 0.025
dex for the CGM and 0.05 dex for the ICM.  Smaller spacing leads to
slightly higher ${\mathcal S}_{\rm stream}$, because turbulent motions
are captured above this scale.  Our method does not work well if there
exist significant deviations from sphericity (e.g. merging galaxies,
massive satellites), or if there are other forces at play that are not
explicitly treated in Equation \ref{equ:effmass}, such as feedback and
viscous forces.  Our tests find robust results beyond 30 kpc at {\it
  M5.3} resolution and 60 kpc at {\it M6.2} resolution.

\subsection{Examples}

\begin{figure*}
\includegraphics[width=0.49\textwidth]{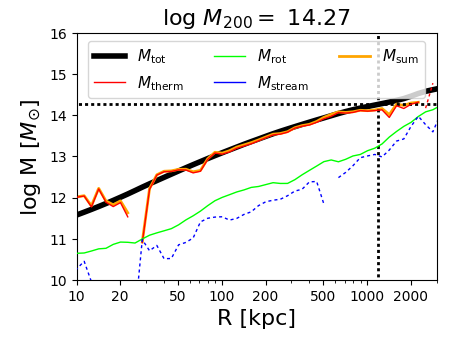}
\includegraphics[width=0.49\textwidth]{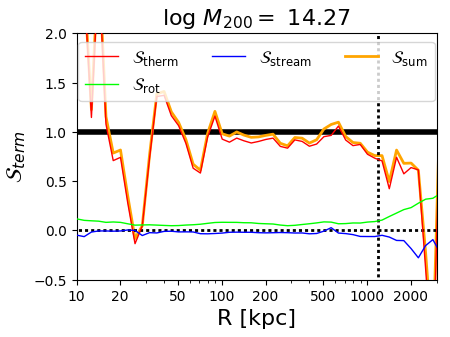}
\includegraphics[width=0.49\textwidth]{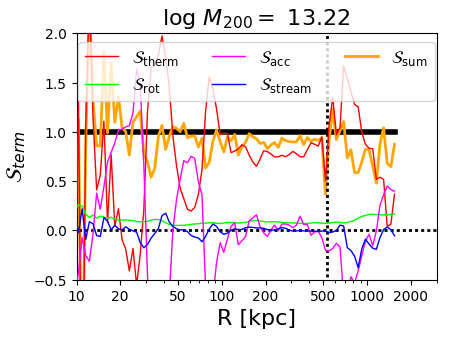}
\includegraphics[width=0.49\textwidth]{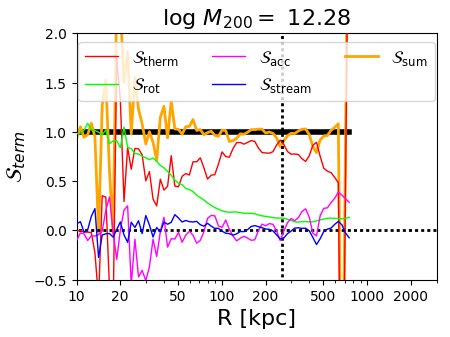}
\caption[]{The effective mass terms in the Euler equation (Equation
  \ref{equ:effmass}) are plotted for a $M_{200}=10^{14.27} \msolar$
  cluster as a function of radius in the upper left.  The thick black
  line is the total gravitational mass ($M_{\rm tot}$), the red line
  represents the thermal pressure gradient ($M_{\rm therm}$), the
  green line represents tangential velocity support ($M_{\rm rot}$),
  the blue line represents the streaming velocity contribution
  ($M_{\rm stream}$), and the orange line indicates the sum of
  $M_{\rm therm}+M_{\rm rot}+M_{\rm stream}$ ($M_{\rm sum}$).
  Coloured dotted lines indicate negative effective mass terms.  The
  vertical dotted black line indicates $R_{200}$ and the horizontal
  black line indicates $M_{200}$.  The upper right plot shows
  normalized ${\mathcal S}$ support terms of the cluster (taking the
  effective mass terms and dividing by $M_{\rm tot}(R)$).  The lower
  two panels show the support terms for a $10^{13.22} \msolar$ group
  halo and a $10^{12.28} \msolar$ $L^*$ halo.  We are able to plot the
  acceleration support (${\mathcal S}_{\rm acc}$) in magenta for these
  haloes and add it to ${\mathcal S}_{\rm sum}$.  Deviations from HSE
  are most significant in the $L^*$ halo, with rotational support
  dominating inside 50 kpc.  Accelerations driven by feedback events
  can be significant in the interiors of group haloes and appear
  necessary for ${\mathcal S}_{\rm sum} \simeq 1$.  We suspect that
  the deviation of ${\mathcal S}_{\rm sum}$ from $1$ for the cluster
  would be remedied by adding ${\mathcal S}_{\rm acc}$, which was not
  available for these haloes.} 
\label{fig:hse_examples}
\end{figure*}

The effective mass terms are shown as a function of radius in the
upper left panel of Figure \ref{fig:hse_examples} for a typical
cluster in our sample with $M_{200}=10^{14.27}\msolar$.  This halo
reproduces similar results reported for clusters \citep[e.g.][]{sut13,
  lau13, nel14}: thermal pressure (red line) balances the
gravitational mass force (thick black line) at the $\sim 90\%$ level,
tangential support (green line) contributes at the $\sim 10\%$ level,
and streaming motions (blue lines) at the several percent level are
more often negative mass terms indicating gas slowing down as it
infalls \citep[cf. ][Figure 5]{sut13}.  Dotted lines for $M_{\rm
  stream}$ indicate negative effective mass values in this panel, but
we plot support terms, $\mathcal S$, on a linear scale
(i.e. normalized to $M_{\rm tot}(R)$) in the upper right panel.
${\mathcal S}_{\rm acc}$ is not saved in our cluster sample, and we
suspect that much of the deviation, especially at small $R$, is the
result of accelerations, likely arising from shocks developed by AGN
superwind feedback.  

The lower two panels show a group halo, $M_{200}=10^{13.22} \msolar$,
and an $L^*$ halo, $M_{200}=10^{12.28} \msolar$, at $z=0$.  The group
shows large variations in ${\mathcal S}_{\rm therm}$ at $<0.2
R_{200}$, which are counterbalanced by ${\mathcal S}_{\rm acc}$.
AGN-driven shocks are propagating through the inner CGM, driven by
discrete black hole accretion episodes in the last 100 Myr.  Crossing
a shock from outside inward, the cooler gas streams inward toward the
shock, the shock front forms a large jump in $\frac{\partial
  P}{\partial r}$ and a large acceleration outward (${\mathcal S}_{\rm
  therm}>1$ and negative ${\mathcal S}_{\rm acc}$).  Then inside the
shock-heated gas tends toward isobaric conditions, which leads to low
${\mathcal S}_{\rm therm}$ and high ${\mathcal S}_{\rm acc}$.  This
rather extreme case at $z=0.01$ conceptualizes how
Equ. \ref{equ:effmass} works, but this halo returns mainly to HSE by
$z=0$ as the sound crossing time of the inner region is less than the
$145$ Myr time interval between the snapshots.

As we move to the $L^*$ halo, we see dramatic differences.  Thermal
support declines to 40-80\% over the majority of the CGM, and
${\mathcal S}_{\rm rot}$ is the largest non-thermal support term
growing steadily toward the center and reaching 70\% at $\sim 30$ kpc.
${\mathcal S}_{\rm acc}$ is negative and ${\mathcal S}_{\rm stream}$
is positive in the inner 70 kpc, which are both indicative of
deviations driven by feedback.  Beyond this radius, radial motion
terms indicate both accretion and outflows.  As in the group halo,
fluctuations in ${\mathcal S}_{\rm acc}$ mirror fluctuations in
${\mathcal S}_{\rm therm}$, indicating accelerations often arise from
deviations in the thermal pressure gradient.  We now move from
individual objects, all of which can be found on our website
http://www.colorado.edu/casa/hydrohalos, to consider the general
trends of our halo samples.

\section{General trends} \label{sec:gentrend}

\subsection{Euler terms}

\begin{figure*}
\includegraphics[width=0.49\textwidth]{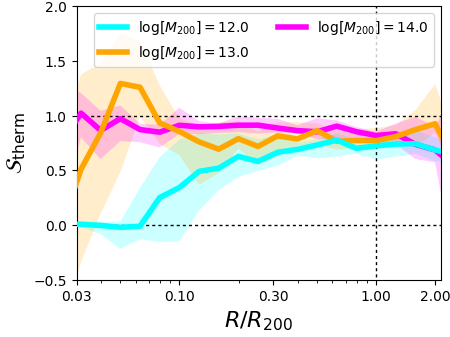}
\includegraphics[width=0.49\textwidth]{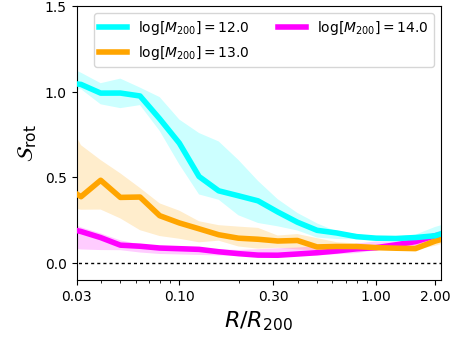}
\includegraphics[width=0.49\textwidth]{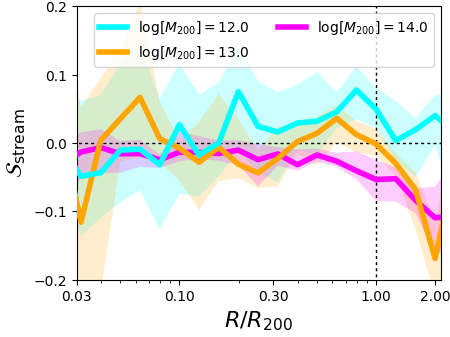}
\includegraphics[width=0.49\textwidth]{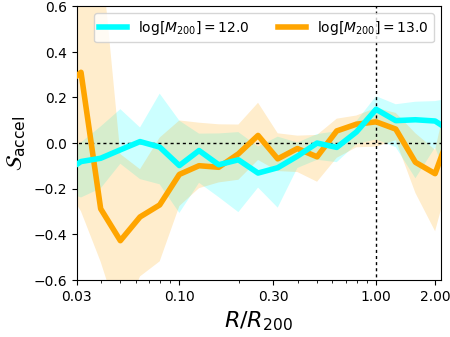}
\caption[]{Euler equation support terms as a function of fractional
  virial radius for our three samples ($L^*$ haloes- aqua, group
  haloes- orange, clusters- magenta).  Median values are shown in
  thick lines and 25-75\% spreads are shaded.
  ${\mathcal S}_{\rm therm}$ dominates (upper left), except in the
  interiors of $L^*$ haloes, where ${\mathcal S}_{\rm rot}$ becomes
  large (upper right).  ${\mathcal S}_{\rm stream}$ and
  ${\mathcal S}_{\rm acc}$ are often non-zero in the interiors of
  $L^*$ and group haloes (lower panels) owing to feedback.
  ${\mathcal S}_{\rm acc}$ is not available for our cluster sample.}
\label{fig:hse_coll}
\end{figure*}

The general trends for the Euler terms appear in Figure
\ref{fig:hse_coll} where we plot the median and 25\%-75\% spread for
our three halo samples: $L^*$ (aqua), group (orange), and cluster
(magenta).  From upper left to lower right, the panels show
${\mathcal S}_{\rm therm}$, ${\mathcal S}_{\rm rot}$,
${\mathcal S}_{\rm stream}$, and ${\mathcal S}_{\rm acc}$ as a
function of fractional virial radius.

Clusters are the most thermally supported, followed closely by groups,
and $L^*$ haloes, which reproduce the trends in the example halo in
Fig. \ref{fig:hse_examples}.  For $L^*$ haloes, thermal support
averages 70-80\% from $R=0.3-1.0 R_{200}$, 40-70\% from $0.1-0.3
R_{200}$.  ${\mathcal S}_{\rm therm}$ drops even further inside $0.1
R_{200}$, although our method loses accuracy here for $L^*$ haloes,
because the stellar and gas discs of similar size break sphericity
plus energy deposition by feedback complicates the analysis.

The upper right panel of Fig. \ref{fig:hse_coll} shows the next
largest contributor, ${\mathcal S}_{\rm rot}$, which includes all
tangential motion, mean and random.  Clusters hold rather steady at
10\%, and groups have more tangential support, averaging 20\% at $0.1
R_{200}$.  On the other hand, $L^*$ haloes show steadily rising
tangential support reaching 30\% at $0.3 R_{200}$ and over 60\% at
$0.1 R_{200}$.  The tangential term for haloes grows toward lower
mass, and can dominate the forces supporting $L^*$ haloes according to
our simulations.

The streaming term includes correlated advection and random motions
above our grid resolution.  The medians of both clusters and groups
are small, but groups show more dispersion among the two samples
indicating more radial motions.  $L^*$ haloes show net positive
streaming, adding 5-10\% to the forces balancing gravity.  We argue
that this term results primarily from outflowing hot gas that loses
velocity at larger radius.  Still, this term is sub-dominant compared
to ${\mathcal S}_{\rm rot}$.  The inertial terms, ${\mathcal S}_{\rm
  rot}$ and ${\mathcal S}_{\rm stream}$, are not invariant under a
Galilean transformation, and become spuriously high if the incorrect
position and velocity are used.  Fortunately, ${\mathcal S}_{\rm rot}$
converges to low values at $R\ga R_{200}$ among the three samples and
shows little dispersion, indicating we have centered our haloes
correctly.  Additionally, velocity profiles explored in Figure
\ref{fig:vel_coll} converge to the expected values at large radii.

Lastly, we show ${\mathcal S}_{\rm acc}$ for just the group and $L^*$
samples. Both groups and $L^*$ haloes more often show negative
acceleration support terms in the interior owing to feedback.
Feedback-driven shocks appear to regularly accelerate gas in the very
interiors of groups at $\la 0.1 R_{200}$.  Gas regularly accelerates
inward between $1-2 R_{200}$ around $L^*$ galaxies indicating
accretion onto the halo.  We do not show median ${\mathcal S}_{\rm
  sum}$ values, but note that the median values are almost always
within $10\%$ of unity as expected.  More often the ${\mathcal S}_{\rm
  sum}$ is lower than one in the CGM, which could be related to the
support from viscous pressure at the locations of shocks, which is not
included in Equ. \ref{equ:effmass} but is implemented in the {\tt
  Anarchy} SPH equation of motion using the \citet{cul10} switch.

\subsection{Velocities} \label{sec:vels}

\begin{figure*}
\includegraphics[width=0.49\textwidth]{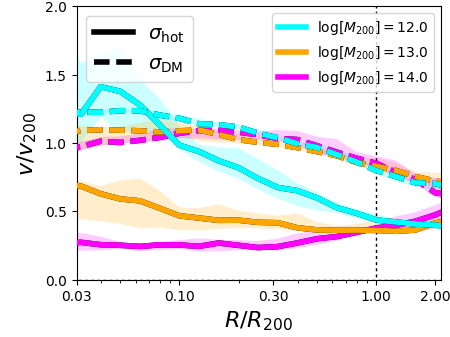}
\includegraphics[width=0.49\textwidth]{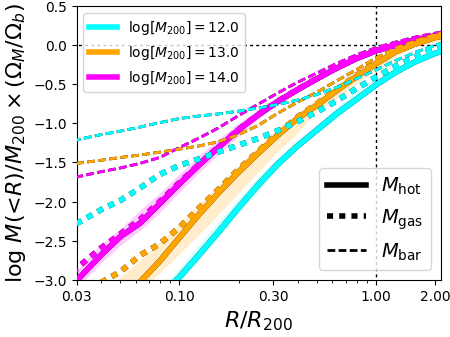}
\includegraphics[width=0.49\textwidth]{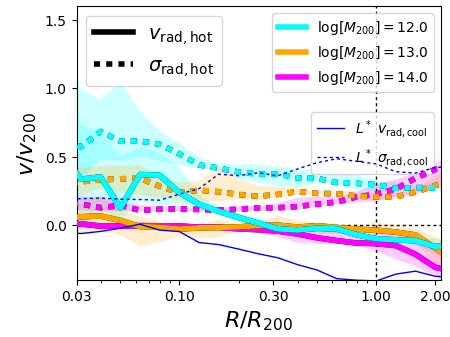}
\includegraphics[width=0.49\textwidth]{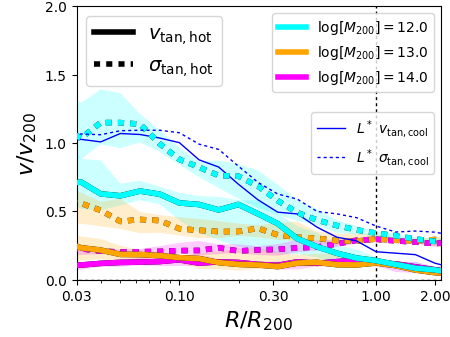}
\caption[]{Median velocities and spreads for our halo samples, as well
  as the cumulative mass plot (upper right panel) in hot gas (solid
  lines), total gas (dotted lines), and all baryons (dashed lines),
  all as functions of fractional virial radius.  The upper left panel
  shows total rms velocity of hot gas only (solid lines) and dark
  matter (dashed lines) normalized to the virial velocity.  The lower
  panels show radial and tangential mean velocities (solid lines) and
  rms velocities (dotted lines) for the hot gas.  Additionally, we
  show the cool gas medians for the $L^*$ sample using thin blue lines
  (solid- mean velocity, dotted- rms velocity) to contrast with hot
  gas.  The cumulative mass plot is normalized to the cosmically
  expected baryonic mass of haloes (horizontal dotted line).  Hot gas
  is kinematically distinct from cool gas, and has significant
  tangential velocities that are part sub-centrifugal rotation and
  part random motions.}
\label{fig:vel_coll}
\end{figure*}

Much of the deviation from HSE owes to motion, which is why we
consider gas velocities in more detail in Figure \ref{fig:vel_coll}.
The upper left panel begins by showing the radially binned
3-dimensional root mean squared (rms) velocity dispersion after
subtracting off the central velocity of the halo, defined as

\begin{equation}
\sigma_k = \sqrt{\frac{\sum_i^n v_{k,i}^2}{n_{k}}},
\end{equation}

\noindent summing over particle indices $i$ for component $k$
(i.e. DM, hot gas, cool gas) in each radial bin containing $n$
particles.  We show the hot gas ($T\geq 10^5$ K, solid lines) and the
dark matter (dashed lines) dispersions for our three halo samples, all
normalized the virial velocity ($v_{200}\equiv \sqrt{G
  M_{200}/R_{200}}$).  The dark matter provides a sanity check,
reproducing the results seen in previous studies
\citep[e.g.][]{nav96}: $\sigma_{\rm DM} \sim v_{200}$, and lower mass
haloes have higher $\sigma_{\rm DM}$ in the interior due to higher
concentrations.

We concentrate on hot baryons, because these are believed to provide
the primary support of the CGM, and we want to explore these motions
specifically to see how the hot gas contributes to the inertial terms
of the Euler equation.  The hot gas has increasing rms velocities for
lower mass haloes that grow larger at lower $R/R_{200}$.  $L^*$ haloes
have hot velocity motions in excess of $v_{200}$ inside $0.1 R_{200}$,
but this is a small fraction of the overall baryons and hot gas does
not dominate the baryon budget until $\sim 0.5 R_{200}$ in this
sample.  The upper right panel shows cumulative baryon profiles for
all baryons (including stars, dashed lines), gaseous baryons (dotted
lines), and hot baryons (solid lines) normalized the cosmically
expected baryonic mass of haloes, $M_{200,{\rm bar}}\equiv
M_{200}\times(\Omega_b/\Omega_M)$.  Clusters are essentially
baryonically closed and dominated by hot baryons, groups retain 60\%
of $M_{200,{\rm bar}}$ in the hot phase inside $R_{200}$, and this
value becomes 1/3rd for $L^*$ haloes.

We subdivide velocities into radial and tangential components in the
lower two panels of Fig. \ref{fig:vel_coll} corresponding to
${\mathcal S}_{\rm stream}$ and ${\mathcal S}_{\rm rot}$.  In each
case, we plot the mean net velocity as a function of $R$
($v_{\rm rad,hot}$ and $v_{\rm tan, hot}$), where
\begin{equation}
v_{\rm rad} = \frac{ \sum_i {\bm v}_i\cdot {\bm R}_i}{\sum_i R_i},  
\end{equation}
and
\begin{equation}
v_{\rm tan} = \frac{\| \sum_i {\bm v}_i\times {\bm R}_i \|}{\sum_i R_i}, 
\end{equation}
\noindent as well as the mean rms velocity dispersion as a function of
$R$ ($\sigma_{\rm rad, hot}$ and $\sigma_{\rm tan, hot}$).

Hot gas shows insignificant net radial inflows in groups and clusters,
although clusters show the largest inflows beyond $R_{200}$ as these
haloes undergo more late-time assembly.  This is consistent with
${\mathcal S}_{\rm stream}$ trending negative outside $R_{200}$.  Hot
radial outflows appear in the interiors of $L^*$ haloes and slow as
they progress outward, and is why we argue hot outflows drive the
positive values of ${\mathcal S}_{\rm stream}$ in
Fig. \ref{fig:hse_coll}.  In contrast the cool, $T<10^5$ K gas shows
net infall, which we show as corresponding thin blue lines only for
the $L^*$ sample.  \citet{van12b} showed that cool gas inflows faster
than hot gas in OWLS simulations, but the thermal feedback
prescriptions in EAGLE result in a net outflow of hot gas.

Hot tangential velocities increase relative to $v_{200}$ from clusters
down to $L^*$ haloes, where they achieve $\sigma_{\rm tan,hot} \sim
v_{200}$ inside $0.1 R_{200}$ indicating a tangential velocity
dispersion capable of supporting the inner halo at $\la 0.1 R_{200}$.
Comparing this to the correlated motion, which is $v_{\rm tan,hot} \ga
0.5 v_{200}$ inside $0.25 R_{200}$, shows the hot halo has significant
yet sub-centrifugal rotation.  It is worth contrasting this to the
cool $L^*$ halo gas, which shows a similar profile for tangential
dispersion as the hot gas, $\sigma_{\rm tan,cool} \sim \sigma_{\rm
  tan,hot}$.  However, $v_{\rm tan,cool} \sim \sigma_{\rm tan,cool}
\sim v_{200}$ (cf. thin blue lines) inside and just beyond $0.1
R_{200}$, indicating fully centrifugally-supported rotating cool discs
extending into the CGM (as opposed to a partially
centrifugally-supported inner hot halo).  To summarize, we argue for
tangential motions providing partial centrifugal support to the inner
hot haloes of $L^*$ galaxies, and discuss the implications of this in
\S\ref{sec:disc}.


\subsection{Masses}

Returning to the upper right panel of Fig. \ref{fig:vel_coll}, a model
hot halo profile of the MW should likely contain less than half
$M_{200,{\rm bar}}$ inside $R_{200}$ despite only $\sim 20-30\%$ of
baryons being accounted for in stellar phase \citep[e.g.][]{mcm11,
  put12}.  This is because simulations with feedback eject a
significant fraction of baryons beyond $R_{200}$ ($\sim 50\%$ for
these EAGLE $L^*$ haloes).  Additionally, a fractionally significant
component of baryons is in the cool phase around these haloes as
calculated most recently by \citet{kee17} and \citet{pro17} from UV
absorption lines, and modeled in detail using these same EAGLE zoom
haloes by \citet{opp17b}.

The physical details of hot halo profiles in EAGLE will be presented
in Davies et al. (in prep.) with results that are promising for X-ray
observational constraints \citep[e.g.][]{and15} at $L^*$ and group
masses.  We also note here that our $L^*$ hot haloes, which account
for a third of $M_{200,\rm bar}$, have very few hot baryons at low
radii. Our MW sample predicts a MW mass halo would have $\sim
2-3\times 10^9 \msolar$ of hot gas inside 50 kpc, which is very close
to the value derived by \citet{mil15} of $3.8\times 10^9 \msolar$ from
$\OVII$ and $\OVIII$ emission lines through the MW halo.

\subsection{Angular momenta}

\begin{figure}
\includegraphics[width=0.49\textwidth]{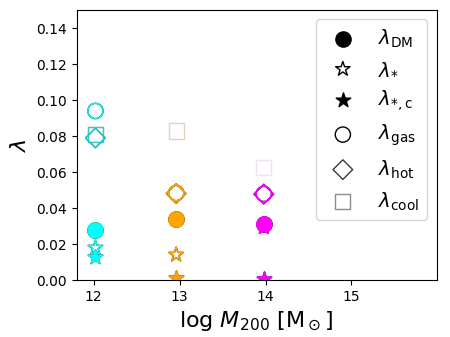}
\caption[]{Median spin parameters of different components (DM- filled
  circles, open stars- all stars in halo, filled stars- stars in
  central only, open circles- all gas, open diamonds- hot gas, \& open
  squares- cool gas) for the 3 halo samples.  Dark matter spin is at
  typical values, $\lambda_{\rm DM}\approx 0.03-0.04$, while gas has
  higher spin parameters around $L^*$ galaxies.  Hot gas dominates the
  angular momentum gas budget in groups and clusters, but approaches
  the dark matter spin at the highest masses.  The cool gas spin
  parameters have shading scaled to the cool CGM gas fractions to
  de-emphasize their importance in massive haloes.}
\label{fig:lambda_coll}
\end{figure}

Given that there is significant correlated rotation in the CGM, we
consider the angular momentum of the CGM and compare it to the other
halo components.  Significant angular momentum in the cool phase is
universal across a range of simulations \citep[e.g.][]{stew17}, which
exceeds the angular momentum of the DM and stars when quantifying the
halo spin parameter,
\begin{equation}
\lambda_k = \frac{j_k}{\sqrt{2} R_{200} v_{200}}
\end{equation}
\noindent using the \citet{bul01} definition for component $k$, where
$j_k$ is the halo-averaged specific angular momentum,
\begin{equation}
j_k = \frac{\| {\bm J}_{k} \|}{\sum_i m_{k,i}}
\end{equation}
\noindent and ${\bm J}_{k}$ is the angular momentum vector,
\begin{equation}
{\bm J}_k = \sum_i m_{k,i} {\bm v}_{k,i}\times {\bm R}_{k,i}
\end{equation}
\noindent summing over particle indices $i$ of component $k$ with
masses $m_{k,i}$.  We plot the halo spin parameters in Figure
\ref{fig:lambda_coll} for different components, including dark matter,
which shows typical values of $\lambda_{\rm DM} \approx 0.03-0.04$ for
our three samples \citep{bul01}.  Stellar spin parameters are less
than $\lambda_{\rm DM}$, although we exclude satellites to focus on
the central galaxy ($\lambda_{*,{\rm c}}$) since including all stars
($\lambda_*$) results in $\lambda_* \approx \lambda_{\rm DM}$ for
clusters.

A number of works have focused on how angular momentum from the cool
CGM translates into the galaxy assembly and morphology
\citep[e.g.][]{dan15, tek15, gar17}.  Usually the cool CGM forms an
extended disc with $\lambda_{\rm cool}$ at least $2\times\lambda_{\rm
  DM}$ around spiral galaxies, which is also seen in our samples.  In
groups and clusters, the fractionally small mass of the cool gas
(shading of $\lambda_{\rm cool}$ scales with cool CGM gas fraction) is
also higher than $\lambda_{\rm DM}$ although it is less clear what the
origin of this angular momentum is.  Excluding satellite ISM gas does
not appreciably change $\lambda_{\rm cool}$, indicating the cool CGM
in massive haloes has the highest angular momentum of any component.  

Considering the hot gas, we expect and find high values of
$\lambda_{\rm hot}$ around our $L^*$ sample compared to groups and
clusters, because of the $L^*$ galaxies' significant hot halo
rotations (\S\ref{sec:vels}).  Indeed $\lambda_{\rm hot}$ is 0.08 for
our $L^*$ sample, or about $3\times$ the spin of the dark matter for
these galaxies, $\lambda_{\rm DM}=0.027$.  Our MW sample has slightly
higher spins: $\lambda_{\rm hot}=0.10$ and $\lambda_{\rm DM}=0.04$,
which are values more representative of the typical $L^*$ halo
selected from a periodic volume.  For the MW sample, the magnitude of
the angular momentum inside $R_{200}$ is the greatest for the hot
halo, $J_{\rm hot} = 10^{14.5} \msolar \kms {\rm kpc}$, compared to
the cool gas, $J_{\rm cool} = 10^{14.0} \msolar \kms {\rm kpc}$, and
stars, $J_{*}=10^{13.2} \msolar \kms {\rm kpc}$.

Of particular interest is the evidence for the Milky Way having a
rotating hot halo from $\OVII$ absorption line profiles, which
\citet{hod16} observed to be spinning at $v=183\pm41 \kms$.  They
calculate $J_{\rm hot}\approx 10^{13.7} \msolar \kms {\rm kpc}$ at 75
kpc, which is equivalent to the amount of angular momentum calculated
to be in the stellar plus $\HI$ MW disc.  Our MW sample finds that
$J_{\rm hot}\approx 10^{13.7} \msolar \kms {\rm kpc}$ at 90 kpc ($0.4
R_{200}$), where it also equals the sum of angular momentum contained
in stars and cool gas.  We discuss the implications of hot rotating
haloes in \S\ref{sec:disc}, where we suggest that more of the
tangential motions of the MW hot halo are in correlated rotation than
the typical galaxy.

\subsection{Energies} \label{sec:energies}

\begin{figure}
\includegraphics[width=0.49\textwidth]{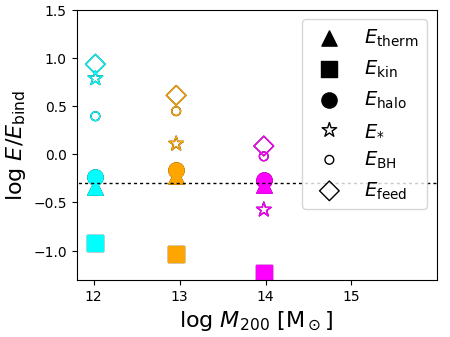}
\caption[]{Median total energies of halo properties, divided by the
  binding energy of the halo.  The total gas thermal energy (filled
  triangles) and kinetic energy (filled squares) sum to define the
  total energy in gas (filled circles), which should be half the
  binding energy of gas for a virialized halo (indicated by the dotted
  line).  The integrated energy from stellar feedback (open stars) and
  BH feedback (small open circles) sum together for the total
  feedback energy (open diamonds).  Feedback energy far exceeds the
  binding energy of $L^*$ and group gaseous haloes, which results in
  significantly rearranged baryonic halo profiles.  This is not the
  case for clusters, which retain their baryons and are closer to
  HSE.}
\label{fig:E_coll}
\end{figure}

The larger velocity components of $L^*$ haloes could indicate a
different partition of energies between thermal and kinetic
components.  However, we find a very mild trend across our samples in
Figure \ref{fig:E_coll}, where we plot energies relative to the
binding energy of the gaseous halo, 
\begin{equation}
E_{\rm bind}= \sum_i \frac{G m_{i} M_{\rm tot}(<R_{i})}{R_{i}} 
\end{equation}
\noindent where we loop over gaseous particle indices $i$ with mass
$m$ and radius $R$. The thermal energy is
\begin{equation}
E_{\rm therm} = \sum_i \frac{3}{2} m_{i} \frac{k_B T_{i}}{\mu m_p}
\end{equation}
\noindent where $k_B$ is the Boltzmann constant, and $T$ is the
particle temperature, $\mu$ is mean molecular weight, and $m_p$ is
proton mass.  The kinetic energy is
\begin{equation}
E_{\rm kin} = \sum_i \frac{1}{2} m_{i} v_{i}^2.
\end{equation}
\noindent $E_{\rm therm}$ dominates ranging from
$3.5\times E_{\rm kin}$ for $L^*$ to $8\times$ for the cluster sample.
$E_{\rm halo} \equiv E_{\rm therm}+ E_{\rm kin}$ is
$\approx E_{\rm bin}/2$ as expected from the virial theorem, and
demonstrates our haloes are in virial equilibrium with gaseous motions
mainly thermalized inside $R_{200}$.  This supports the finding of
\citet{opp16} using these same simulations to show that high oxygen
ions, which are collisionally ionized, trace the virial temperatures
of haloes, $T_{\rm vir} \sim M_{200}^{2/3}$.  $\OVI$ is argued to
appear around $L^*$ and not group haloes, because the former have
temperatures that overlap the $T\approx 10^{5.5}$ K $\OVI$
collisionally ionized temperature.  Despite the values of
$\sigma_{\rm hot}$ in $L^*$ haloes exceeding $v_{200}$ at
$R<0.1 R_{200}$, the kinetic energy contribution is comparatively
small, since most of the hot gas resides at larger radii for these
haloes.

The source of the velocities are in part related to the process of
feedback, which we quantify as $E_{\rm feed} \equiv E_{*} + E_{\rm
  BH}$, where stellar feedback is
\begin{equation}
E_* = \epsilon_{\rm SF} M_*,
\end{equation}
\noindent and feedback from BH accretion is
\begin{equation}
E_{\rm BH} = \epsilon_{\rm BH} M_{\rm BH}.
\end{equation}
\noindent Stellar feedback efficiency using the EAGLE Recal
prescription is $\epsilon_{\rm SF}\approx 1.75\times 10^{49}$ erg
$\msolar^{-1}$ \citep{cra15}, and we add a multiple of 1.8 to present-day
stellar masses to account for stellar death \citep{opp16}.  The EAGLE
$\epsilon_{\rm BH}$ is 1.5\% of the rest-mass energy of the black hole
\citep{sch15}.  Feedback energies are renormalized down by the mass of
baryons inside $R_{200}$ divided by $M_{200,{\rm
    bar}}$.\footnote{Feedback energies should be compared to the
  binding energy of all baryons that would be inside $R_{200}$ without
  feedback, and this provides a correction to these energies ($E_*$,
  $E_{\rm BH}$, \& $E_{\rm feed}$) of $0.5$, $0.7$, and $0.9$ relative
  to $E_{\rm bind}$ for $L^*$, group, and cluster haloes,
  respectively.}  Both forms of feedback are imparted thermally, which
likely increases the hot gas motions relative to a kinetic feedback
prescription.  However, the feedback behaves very differently at the
threshold mass of $M_{200}\approx 10^{12}\msolar$.  \citet{bow17} used
EAGLE to demonstrate stellar-driven thermal feedback can rise
buoyantly through the ambient halo medium below $10^{12} \msolar$, but
becomes ineffective above $10^{12} \msolar$ as the adiabat of the
heated gas no longer exceeds that of the inner CGM.  We argue that the
hot halo motions for the $10^{12} \msolar$ $L^*$ galaxies arise from
stellar feedback that rose buoyantly through the halo many Gyrs
earlier when the halo was less massive and had lower pressure.  This
feedback promotes low-entropy, low-angular momentum gas from the
center of the galaxy \citep[e.g.][]{gov10,bro12,ubl14,chr16} via
buoyant adiabatic expansion to radii $\ga R_{200}$.  The $\sim
10^{5.5}$ K $\OVI$ haloes in these zooms \citep{opp16} are a remnant
of this process.  The feedback launched at an average $z\approx 1$ can
in part explain the strong $\OVI$ observed by COS-Halos at impacts up
to 150 kpc from $L^*$ galaxies \citep{tum11}, but residing at
$R=150-500$ kpc according to our zooms.  Some of this hot gas at $R\ga
R_{200}$ journeys back into the inner halo, having gained angular
momentum, but not necessarily in a coherent direction.  Therefore, the
mean tangential motions are significantly larger than the correlated
rotation.  These trends were seen in $L^*$ haloes in EAGLE volumes by
\citet{ste17}, who also reported significant hot halo angular momentum
with a spin direction that is usually offset from the cooling gas that
builds the disc.  Hence, it is not surprising to find significant
tangential motion in both hot and cool components of the CGM that is
not necessarily correlated in direction and velocity
(Fig. \ref{fig:vel_coll}, lower panels, cf. cyan and blue lines).

The energy imparted around group haloes is dominated by BH feedback,
which acts in a preventative manner by heating halo gas.  The buoyant
launching method is ineffective here \citep{bow17}, so despite $E_{\rm
  feed} \gg E_{\rm bind}$, group haloes do not eject as much material
beyond $R_{200}$ (Fig. \ref{fig:vel_coll}, upper right).  The angular
momentum of the hot gas is not promoted by AGN-driven shocks that have
a high degree of spherical symmetry around the central galaxy.  These
shocks are a feature of EAGLE feedback that may not be reproduced in
the real Universe where wind-driven bubbles are offset from the galaxy
center, often in bipolar structures \citep[e.g.][]{mcn14}.  Virial
shocks have primarily processed the ICM, resulting in efficient
thermalization, and low velocities (relative to $v_{200}$) in the
interior of our cluster sample.


\section{Discussion} \label{sec:disc}

We have examined the hydrodynamic state of $L^*$, group, and cluster
haloes, and determined that baryons deviate from HSE the most at lower
masses.  We begin our discussion by considering how baryons break the
near self-similarity expected from dark matter halo theory, and how
this manifests itself in the CGM.

\subsection{Breaking the self-similarity of the CGM} \label{sec:selfsim}

Dark matter-only simulations indicate $L^*$ and cluster haloes, while
not completely scale invariant owing to $\Lambda$CDM cosmology, have
very similar structures \citep[e.g.][]{kly99,moo99}.  A cluster
exhibits lower concentration and more infall at the virial radius due
to later assembly, but these effects are small compared to processes
we refer to as ``baryonic processing.''  Baryons added into an
adiabatic simulation (i.e. no cooling) are retained within $R_{200}$
independent of virial mass \citep[][]{cra07}, although there are small
deviations from scale-free density and temperature profiles across
cluster masses \citep{asc06}.  We concentrate on the two largest
sources of baryonic processing, cooling and feedback.

\begin{figure*}
\includegraphics[width=0.49\textwidth]{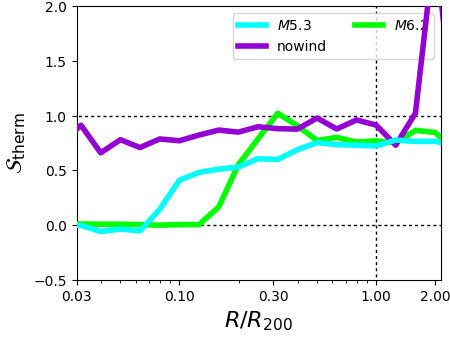}
\includegraphics[width=0.49\textwidth]{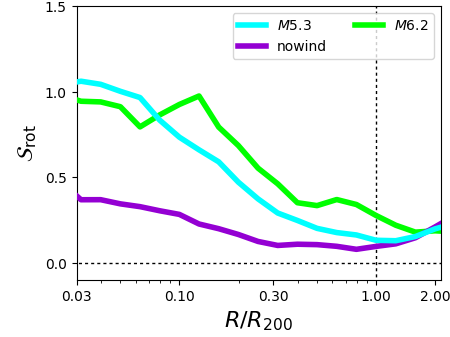}
\includegraphics[width=0.49\textwidth]{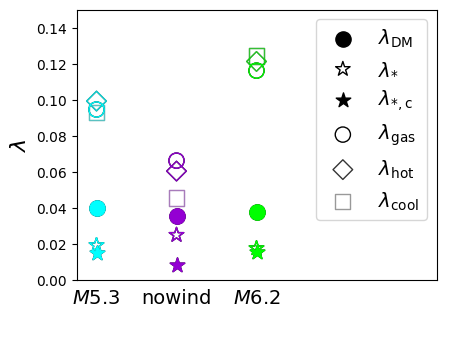}
\includegraphics[width=0.49\textwidth]{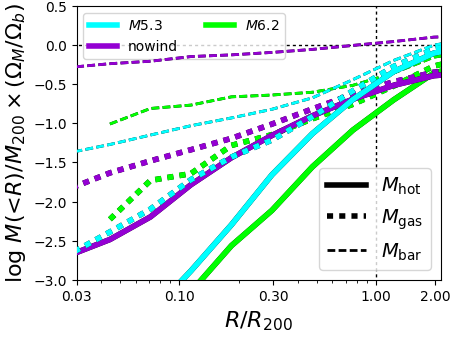}
\caption[]{Median relations of the Milky Way sample at the fiducial
  {\it M5.3} resolution (aqua), for no-wind runs at {\it M5.3}
  resolution (violet), and at {\it M6.2} resolution (green).  We show
  the normalized Euler terms for thermal pressure (upper left) and
  tangential velocity support (upper right).  The spin parameters
  broken down by component are shown in the lower left panel and the
  cumulative mass plot of baryons are shown in the lower right panel.
  We argue that the hot gas in no-wind $L^*$ haloes holds similar
  hydrodynamic properties as the ICM in clusters, and that the primary
  driver of deviations from HSE around $L^*$ and group haloes is
  feedback.  The lower resolution {\it M6.2} $L^*$ haloes exhibit
  similar deviations from HSE as their {\it M5.3} counterparts.}
\label{fig:other_coll}
\end{figure*}

\noindent{\bf Cooling:} To isolate the effect of cooling, we use our 5
halo no-wind MW sample, plotted for several of our median relations in
violet in Figure \ref{fig:other_coll}.  Compared to the 5 halo MW
EAGLE sample (cyan, labeled {\it M5.3}), we see much higher ${\mathcal
  S}_{\rm therm}$ (upper left), much lower ${\mathcal S}_{\rm rot}$
(upper right), and significantly less spin in the hot gas, cool CGM,
and stars in the central galaxy (lower left).  To isolate the effect
of cooling, we compare the no-wind sample to the cluster sample where
gas is heated to such high temperatures that cooling becomes
inefficient.  The hydrodynamic state of the no-wind CGM holds more
similar characteristics with the cluster sample (${\mathcal S}_{\rm
  therm}$ dominates everywhere, ${\mathcal S}_{\rm rot}$ is small,
$\lambda_{\rm hot}$ is smaller). 

The cluster versus no-wind comparison does not purely isolate cooling,
because the former does have stellar and BH feedback.  The median
stellar baryon fraction in stars is 7.8\% in clusters and 79\% in
no-wind zooms, leaving 89\% of baryons in the ICM and 30\% in the
no-wind CGM\footnote{No-wind $L^*$ haloes appear to exceed
  $M_{200,{\rm bar}}$ by 10\% in our accounting.} (cf. lower right
panel of Fig. \ref{fig:other_coll}).  A no-wind cluster simulation
would certainly have a higher stellar baryon fraction ($\sim
20-40\%$), but most of the baryons would remain in the ICM based on
previous studies \citep{lew00}.  However, we argue that the no-wind
$L^*$ haloes with one dominant galaxy results in significant baryonic
contraction \citep{blu86} that exceeds the DM assembly effect of
higher concentrations for $L^*$ haloes, resulting in more rotation and
higher hot gas spin parameters.  Nevertheless, despite this effect,
the hydrodynamic states of the no-wind and cluster samples are very
similar considering the gas from $0.3-1.0 R_{200}$.  Thermal support
dominates in similar fashion, despite the fact that clusters retain
$2/3$rd of $M_{200,{\rm bar}}$ between these radii and no-wind $L^*$
haloes sum to only 18\% of $M_{200,{\rm bar}}$ over the same radii.

\noindent{\bf Feedback:} We therefore put forth that feedback is the
primary driver for deviations from HSE in the CGM.  Net acceleration
out of the galaxy and significant dispersions in streaming suggest
decelerating outflows contribute to larger deviations from HSE for
$L^*$ galaxies compared to their no-wind analogues.  These ongoing,
transient processes contribute to the continual and cumulative
build-up of tangential motions supporting the inner CGM.  In $L^*$
haloes, buoyant thermal winds travel to the outer CGM \citep{bow17},
gaining angular momentum as they re-accrete back into the inner CGM
and onto the galaxy.

Other simulations find that winds leaving the disc for long times,
($>1$ Gyr), travel large distances ($50-100$ kpc), and always gain
angular momentum before re-accreting onto the galaxy \citep{ubl14}.
\citet{chr16} showed that gas recycled onto galaxies is ``spun-up'' by
the halo and gains $2-3\times$ more angular momentum than it had
before it was ejected.  However, few simulations have considered the
angular momentum of the hot halo gas itself.  Our work does not have
hot haloes spinning at the virial velocity, otherwise the halo would
arrange itself in a kinematic hot disc.  Our haloes are essentially
spherical, having significant uncorrelated tangential motion that must
be continually replenished by feedback otherwise such motions would
dissipate in a spherical geometry.  The work of \citet{ste17} buoys
our results, quantifying the significant angular momentum of the hot
CGM in EAGLE haloes, but finding its axis usually misaligned with the
cool gas disc such that gas cooling out of the hot halo precesses to
align with the cool gas disc.  Hence, rotating hot haloes may not
necessarily align with stellar discs, and the significant magnitude of
the tangential motion in the inner halo is not all in a correlated
spinning halo (cf. cyan dotted and solid lines in lower right panel of
Fig. \ref{fig:vel_coll}).

Fig. \ref{fig:other_coll} also shows the MW halo sample for low
resolution ({\it M6.2}- green), which we discuss in the Appendix
\ref{sec:res}.  We summarize here that our main results are found
across different resolutions, but there are significant differences in
the hot CGM content within $R_{200}$.

\subsection{Observational evidence and prospects for rotating hot haloes}  \label{sec:obs}

The clearest evidence for rotating hot haloes may be our own Milky
Way, which shows evidence of a sub-centrifugal rotating hot halo at
$183\pm 41 \kms$, or about $3/4$th the MW disc rotation speed
\citep{hod16}.  This supports our findings of significant tangential
motion in our primary $L^*$ zooms.  While the MW shows more coherent
rotation that appears aligned with the disc compared to EAGLE
galaxies, it may be linked to the relatively unperturbed nature of the
Milky Way stellar disc with little evidence of a major merger in the
last $10$ Gyr \citep{ste08}.


Detecting rotating hot haloes around other galaxies is beyond the
capability of present-day X-ray telescopes, but instrument technology
capable of $R\ga 3000$ resolution to resolve sub-$100 \kms$ $\OVII$
and $\OVIII$ line profiles could resolve these velocities
\citep{mil16}.  However, nearer term observational prospects may rely
on observing the gas cooling out of rotating hot gas, and relating
this process to warped galactic disc structures.  \citet{ros10}
examined a high-resolution cosmological zoom simulation finding that
cool, accreting gas is strongly torqued by spinning hot halo gas.
They argued a misaligned warped disc of newly accreted material is
indicative of a hot halo spinning on an unaligned axis.

Linking cooling gas structures in the CGM to morphological disc
structures could be done with current telescopes.  A Cosmic Origins
Spectrograph (COS) survey targeting gas along the semi-major axis of
edge-on disc galaxies could relate $\HI$ and metal absorption
kinematics to these galaxy's low-surface brightness extended
structures indicating the presence or absence of warps, and ultimately
the influence of the rotating hot halo.  Already, tantalizing evidence
exists from the COS-GASS/COS-Halos kinematic compilation of
\citet{borth16} showing that while there is ubiquitous $\HI$ around
star-forming galaxies, it becomes increasingly kinematically offset
from the central galaxy velocity at lower impact parameters.






\subsection{Future directions}  \label{sec:future}

\subsubsection{Observatories}

Future approved and proposed X-ray missions face the challenge of
detecting hot CGM profiles tracing the majority of a halo's baryons,
which have been long known to diverge from self-similar scaling
relations \citep[e.g.][]{whi91, bab02, dav02, cra10}.  While feedback
significantly reduces X-ray luminosities by redistributing hot baryons
into a more extended and diffuse distribution, the velocity structures
of the hot baryons, especially those in the interior CGM, also become
significantly perturbed by feedback.  A far more complete
understanding of the dynamics of the hot CGM requires not just
sensitivity to detect baryons out to $R_{200}$, but the velocity
resolution to observe the hydrodynamics in the inner CGM.  Thus, X-ray
mission concepts, especially NASA's {\it Lynx} large strategic mission
concept \citep{gas16}, should consider how to resolve the velocity
structure within $0.3 R_{200}$ in a survey of Milky Way-like haloes.

The proposed {\it Arcus} NASA Explorer mission \citep{smi16} could
uncover hot gas kinematics around MW-like galaxies early in the next
decade.  With a $R=2500-5000$ resolution grating spectrometer
resolving velocities $<100 \kms$, the hot CGM kinematic spread and
velocity offset from the galaxy's systematic velocity can be
determined via quasar absorption line spectroscopy of $\OVII$ and
$\OVIII$ lines that are expected to be $\ga 10$ m\AA~in strength at
$R\la 100$ kpc.  This spectral resolution combined with a collecting
area of $>400$ cm$^2$ provides an order of magnitude improvement over
{\it Chandra} and {\it XMM} grating spectrometers.

However, sensitivity of emission out to $R_{200}$ to attempt to
ascertain the masses and the thermodynamic states of isolated spiral
and elliptical galaxy haloes requires a high-resolution
microcalorimeter on a mission like {\it Lynx}.  Our extension of the
formalism developed to derive cluster masses and quantify deviations
from HSE could apply to the CGM as the outer CGM beyond $0.3 R_{200}$
is primarily supported by the thermal pressure gradient.  The zooms in
this paper were previously used to argue that COS-Halos passive
galaxies live in haloes $\sim 10\times$ more massive than COS-Halos
star-forming galaxies \citep{opp16}, but only sensitive X-ray
observations can weigh these haloes {\it and} reveal how the
thermodynamic processes of virialization and feedback distribute the
bulk of a halo's baryons.


\subsubsection{Analytic models}

It is crucial to understand the hydrodynamics of the hot phase of the
CGM for how $L^*$ galaxies grow, even though this phase is less
important than the cool phase for fueling star formation
\citep[e.g.][]{ker05, dek06}.  Although the mass of the inner CGM at
$<0.3 R_{200}$ is dominated by cool gas (Fig. \ref{fig:vel_coll},
upper right panel), and this is consistent with a range of low-ion
metal absorption line observations \citep{opp17b}, the hot phase
determines the medium for those cool clouds.  If the inner hot halo is
rapidly rotating and also has significant uncorrelated radial and
tangential motions, then the pressure profile will be different than
the HSE assumption.  This makes static analytic hot halo models like
\citet{mal04} obsolete.  The claim that cool clouds are out of
pressure equilibrium with the hot phase at a level of $100\times$ as
\citet{wer14} holds no relevance for cosmologically-based simulations
that include the effect of significant feedback.  A completely new set
of analytic models that include kinetic haloes is required to
understand the formation, survival, and destruction of cool clouds
that comprise the majority of CGM absorption line measurements
\citep[e.g.][]{sto13, wer13, bor14, lia14, borth15, bur15, joh15}.
Otherwise, one is considering the wrong models for how gas feeds
galaxies.

\citet{pez17} developed analytical models of hot haloes with
sub-centrifugal rotation, which they argued are necessary for the
inside-out growth of disc galaxies using cosmologically motivated
angular momentum distributions.  They motivated a model where
super-virial temperatures in the inner CGM, as indicated by
observations \citep{mil15}, require ejective feedback to remove low
angular momentum gas and place it in the outer halo.  The mixing of
specific angular momentum from cosmological accretion and recycled gas
adds further complexity to models like \citet{pez17}.  Adding in the
processes of gas cooling out of dynamic hot haloes, the survival of
cool filaments under the shear of hot gas rotation, and the
redistribution of feedback-driven gas and its angular momentum are
relevant next steps in such explorations.


\subsubsection{Simulations}

Of course cosmologically-based simulations with refined feedback
prescriptions \citep{hopk14, vog14, sch15, dav16} include all of the
above processes, but their complexity can inhibit easy interpretation.
Our decomposition of the Euler terms in a single dimension (radial)
reveals interesting trends as to what provides support to the CGM, but
a full force ``audit'' of a simulation can reveal the hydrodynamical
interactions setting the force balance at the surfaces between each
resolution element.  What torque does a rotating hot halo have on
structures like the Magellanic stream and high-velocity clouds
\citep[e.g.][]{sal15}?  How do the shear forces between such a hot
halo spinning on an axis misaligned with an extended cool disc
manifest themselves in warped disc structures \citep[e.g.][]{ros10}?
Is collisionally ionized $\OVI$ with cooling times much less than a
Hubble time continually excited by weak shocks from rotational
motions?


Different numerical methods will give different quantifications of
forces active on the multiphase medium, but applying this type of
formalism also allows other forces not included in our simulations to
be quantified.  Non-thermal pressure sources including magnetic fields
and cosmic rays could be significant in other models
\citep[e.g.][]{fae17} and simulations \citep[e.g.][]{sal16, nel17}.
\citet{wer14} and \citet{mcq17} argued that non-thermal pressures in
cool, $T\sim 10^4$ K clouds are needed to balance the thermal pressure
of the hot medium.  The \citet{sal16} simulations show significant
cosmic ray pressure in the cool CGM, while the hot CGM is still
dominated by thermal pressure.


\section{Summary} \label{sec:summary}

We examine the hydrodynamic state of haloes hosting normal galaxies to
determine how the circumgalactic medium deviates from hydrostatic
equilibrium (HSE).  Our study uses simulated clusters as a reference
point, which are confirmed to be in HSE at the $\approx 90\%$ level
\citep[e.g.][]{sut13}.  The CGM in group ($M_{200}\approx 10^{13}
\msolar$) and $L^*$ ($M_{200}\approx 10^{12} \msolar$) haloes are not
scaled down versions of clusters, but have larger deviations from HSE,
especially in their interiors.  Tangential motions contribute the
largest deviation from HSE, especially around $L^*$ galaxies.  The
motions not only include sub-centrifugal rotation of hot halo gas, but
also uncorrelated tangential motions that are continually replenished
by thermal feedback.  Radial streaming motions and acceleration
related to outflows are also significant in $L^*$ haloes, especially
inside $0.3 R_{200}$ where thermally-supported HSE is a poor
description.

Haloes from $M_{200} \approx 10^{12}$ to $\ga 10^{14.5} \msolar$ are
well-described as being in virial equilibrium with thermal energy
dominating over kinetic energy.  Stellar and BH feedback are
primarily responsible for disrupting the expected scaling relations at
the low-mass end of the hot halo regime.  While cooling is also more
efficient for these haloes, the ability of feedback to overcome the
binding energy of the halo gas drives up to half of the baryons beyond
$R_{200}$ and transfers significant angular velocity to re-accreting
gas.  $L^*$ haloes advect low-angular momentum disc gas via buoyant
thermal feedback \citep{bow17} to the outer CGM, from where hot gas
re-accretes over the course of many Gyr spinning up the hot haloes and
providing excess uncorrelated tangential velocities.  Groups also
impart significant feedback energy relative to their binding energies,
but AGN heating acts primarily as late-time preventative feedback
unable to remove baryons from $R_{200}$ while not adding to tangential
velocities.\footnote{Visit http://www.colorado.edu/casa/hydrohalos for
  visualizations of these processes caught in the act.}

The EAGLE simulations used in this investigation have been extensively
tested against observations of both the CGM and galaxies.  Our zooms
reproduce key observations of CGM metal absorption \citep{opp16,
  opp17a, opp17b}, $\HI$ absorption properties (Horton et al., in
prep.), and X-ray emission properties (Davies et al., in prep.).  Our
Milky Way sample masses, velocities, and angular momenta of the inner
hot halo compare well to X-ray-derived values of our halo
\citep{mil15,hod16}, and lend credence that spinning hot haloes are
not just theoretical.  These zooms are part of the broader EAGLE suite
of simulations that make successful predictions for a variety of
galaxy observables, including the galactic stellar mass function, to
which the model was calibrated \citep{sch15}.  Hence, EAGLE
simulations are especially powerful for motivating new observational
techniques focusing on the kinematics of the hot and cool CGM around
normal galaxies.  We therefore advocate observational campaigns that
relate the hot gas motions in the CGM to extended cool gas structures
and morphological stellar disc features as a priority in the
understanding of galaxy assembly.



\section*{acknowledgments}

Valuable discussions with Joel Bregman, Romeel Dav\'e, Neal Katz,
Anatoly Klypin, Gabriele Pezzulli, Mary Putman, John Stocke, and
Gurtina Besla contributed to this work.  In particular, BDO
acknowledges the detailed e-mail discussions with Yasushi Suto about
Euler terms, the discussion of viscosity with Vladimir Zhdankin, and
the input from Ryan Horton enhancing the clarity of the manuscript.
BDO is grateful for Michael Wyner for copyediting the final submitted
manuscript.  The author thanks the anonymous referee for their review
that improved this manuscript.  The author thanks the EAGLE consortium
for providing public access to the simulations.  The Hubble Theory
grant HST-AR-14577 supported this work.  This work used the DiRAC Data
Centric system at Durham University, operated by the Institute for
Computational Cosmology on behalf of the STFC DiRAC HPC Facility
(www.dirac.ac.uk). This equipment was funded by BIS National
E-infrastructure capital grant ST/K00042X/1, STFC capital grants
ST/H008519/1 and ST/K00087X/1, STFC DiRAC Operations grant
ST/K003267/1 and Durham University. DiRAC is part of the National
E-Infrastructure.  Analysis was run at Leiden Observatory with the
support of Erik Deul and David Jansen.

\appendix

\section{Numerical convergence} \label{sec:res}

The {\it M6.2} MW halo sample is also shown in Figure
\ref{fig:other_coll} in green. The {\it M6.2} resolution (equivalent
to Ref-L100N1504) shows $2\times$ weaker $\OVI$ column densities
\citep{opp16}, which was also seen in \citet{rah16}.  There are
significant differences between the resolutions when examining their
physical properties.  As resolution decreases, ${\mathcal S}_{\rm
  rot}$ increases and $\lambda_{\rm hot}$ increases from $2.5$ to
$3\times \lambda_{\rm DM}$.  Note that the gridding is coarser at
lower resolution (\S\ref{sec:sim}), and that the inner radius we trust
for the Euler terms is log $R/R_{200}\approx -1.0$ and $-0.7$ for {\it
  M5.3} and {\it M6.2}, respectively.

The cumulative baryon plot in the lower right of
Fig. \ref{fig:other_coll} demonstrates increasing hot CGM masses
within $R_{200}$ with resolution ($M_{\rm hot}/M_{200,{\rm bar}}=
15\%$ for {\it M6.2} and $35\%$ for {\it M5.3}).  This is a
significant difference, and demonstrates in part the EAGLE ``weak''
resolution convergence strategy of altering the prescription (Ref
vs. Recal) at different resolutions to attempt to reproduce the same
galaxy properties \citep{sch15} can lead to other parameters being
unconverged.  Thermally heating winds with $8\times$ more mass per
feedback event from {\it M5.3} to {\it M6.2} results in gas that
travels further (more often outside $R_{200}$) and achieves greater
angular momentum in the CGM leading to higher ${\mathcal S}_{\rm
  rot}$.  However $\lambda_{\rm hot}$ varies less between resolutions
and is significantly enhanced by feedback no matter the resolution.

Feedback spins up hot haloes at both resolutions, but which is most
correct for the CGM?  We advocate the {\it M5.3} zooms and results
from the Recal-L025N0752 volume, because of their ability to reproduce
a range of CGM observables, including X-ray properties that are
sensitive to the hot baryon content of haloes, which is most obviously
variable across resolutions.  Whether this mass resolution has
physical meaning for the mass scales of feedback-driven buoyant hot
bubbles in the real Universe is still to be determined.

\end{document}